\documentclass[prb,aps,twocolumn,superscriptaddress,showpacs]{revtex4-2} 
\usepackage{graphicx,color,xcolor}
\usepackage{latexsym}
\usepackage{amsmath,amsfonts,amssymb}
\usepackage[colorlinks,linkcolor=blue,urlcolor=blue,citecolor=blue]{hyperref}
\usepackage[all]{hypcap} 
\begin{document}

\newcommand\calR{\mathcal{R}}
\newcommand\msc[1]{\textcolor{red}{\textbf{MSC: #1}}}

\title{Probing Kondo spin fluctuations with scanning tunneling microscopy and electron spin resonance}
\author{Yinan Fang}
\affiliation{Beijing Computational Science Research Center, Beijing 100193, People's Republic of China}
\affiliation{School of Physics and Astronomy, Yunnan University, Kunming 650091, China}

\author{Stefano Chesi}
\email{stefano.chesi@csrc.ac.cn}
\affiliation{Beijing Computational Science Research Center, Beijing 100193, People's Republic of China}
\affiliation{ICTP, Strada Costiera 11, I-34151 Trieste, Italy}
\affiliation{Department of Physics, Beijing Normal University, Beijing 100875, People's Republic of China}

\author{Mahn-Soo Choi}
\email{choims@korea.ac.kr}
\affiliation{Department of Physics, Korea University, Seoul 02841, South Korea}

\date{\today}

\begin{abstract}
We theoretically analyze a state-of-the-art experimental method based on a combination of electron spin resonance and scanning tunneling microscopy (ESR-STM), to directly probe the spin fluctuations in the Kondo effect. The Kondo impurity is exchange coupled to the probe spin, and the ESR-STM setup detects the small level shifts in the probe spin induced by the spin fluctuations of the Kondo impurity. We use the open quantum system approach by regarding the probe spin as the ``system'' and the Kondo impurity spin as the fluctuating ``bath'' to evaluate the resonance line shifts in terms of the dynamic spin susceptibility of the Kondo impurity. We consider various common adatoms on surfaces as possible probe spins and estimate the corresponding level shifts. It is found that the sensitivity is most pronounced for the probe spins with transverse magnetic anisotropy.
\end{abstract}

\maketitle

\section{Introduction}

Ever since the first theoretical explanation, dating back more than 50 years \cite{1964_ProgTheorPhys_Kondo}, the Kondo effect has attracted great
interest not only in condensed matter physics but also in other areas of physics \cite{1984_PRB_Tachiki,2001_RMP_Stewart,2009_RMP_Alloul,1988_JETP_Glazman,2012_PRL_Lee,2013_PRB_Pillet}.
It serves as a physical paradigm to understand other strong correlation
effects and helped the developments of powerful theoretical
methods \cite{1975_RMP_Wilson,1994_JPCondMat_Costi,2008_RMP_Bulla,1980_PRL_Andrei,1984_PRL_Andrei}.
Recent advances in nanotechnology, allowing the control of electronic states down to a single magnetic impurity, have brought renewed
interest in this topic \cite{1998_Nature_Kastner,1998_Science_Cronenwett,
 2000_Science_vanderWiel}, and more exotic Kondo physics has been designed
and explored \cite{2005_PRL_Choi,2006_PRB_Lim,2009_NatPhys_Delattre}.

Despite this progress, the answer to some of the most fundamental questions, such as the size of the
Kondo screening cloud \cite{2001_PRL_Affleck,2008_PRB_Affleck,2013_PRL_Park,2020_Nature_Borzenets} (which is an essential feature of the Kondo effect), 
still remain elusive due to the lack of direct experimental probes. 
One of the measurements closest to a direct assessment of the
Kondo screening cloud considers the compressibility (i.e., the
charge susceptibility) of the Kondo impurity \cite{2017_Nature_Desjardins}. By
measuring the suppression of the compressibility in the presence of a
Kondo-enhanced transport current, the authors of Ref.~\onlinecite{2017_Nature_Desjardins} have demonstrated that the Kondo effect
is due to spin fluctuations without charge fluctuations. However, the experiment probed only the charge sector but not directly the spin fluctuations themselves.


Here, we propose a method to directly probe the spin fluctuations of the Kondo effect by taking advantage of the excellent spatial ($\textrm{\AA}$) and energy ($\mu \textrm{eV}$) resolution \cite{1987_RMP_Binnig, 1998_Science_Madhavan, 2009_RMP_Wiesendanger, 2011_NatPhys_Pruser} of scanning tunneling microscopy (STM) and spectroscopy (STS). Before discussing our scheme, we note how STM and STS have allowed many remarkable experiments, such as the observation of charge density oscillations surrounding a magnetic adatom \cite{1993_Nature_Crommie,1999_JAP_Kawasaka}, and later of the Kondo resonance in Ce or Co adatoms \cite{1998_PRL_Li,1998_Science_Madhavan}.
Varying the lateral tip-adatom distance has revealed links between the asymmetry of the Kondo resonance and different types of tunneling matrix elements \cite{1998_Science_Madhavan,2000_PRL_Ujsaghy}.
The capability of manipulating adatoms, vividly illustrated by the quantum mirage \cite{2000_Nature_Manoharan,2003_RMP_Fiete},
has allowed controllable studies on the effect of interaction between two nearby Kondo impurities \cite{1999_PRB_Chen,2009_PRL_Otte}, such as in the two-impurity Kondo model \cite{2009_PRL_Sela,2011_Nat_Phys_Bork}.
Equipped further with spin-polarized tips, STM can also investigate magnetic properties such as magnetic anisotropy
\cite{2008_Nat_Phys_Otte,2013_Nat_Nano_Oberg} and magnetic exchange coupling \cite{2009_PRL_Otte} at the atomic level.

\begin{figure}
\begin{centering}
\includegraphics[width=8.6cm]{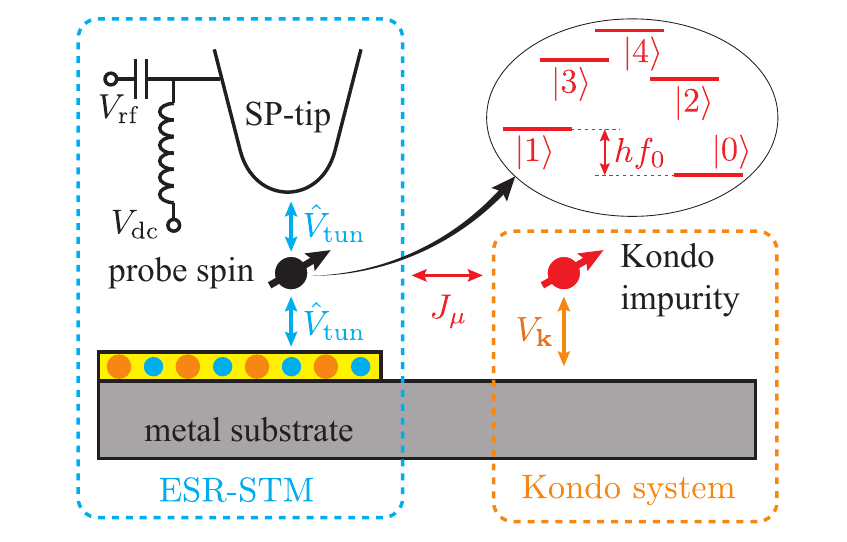}
\par\end{centering}
\caption{Schematics of a Kondo impurity (right dashed box) coupled to
a probe spin measured by an ESR-STM device (left dashed box). The
Kondo impurity is tunnel-coupled with electrons in the metal
substrate, and can be described by the Anderson model \citep{1993_BOOK_Hewson}. 
The probe spin is adsorbed on the top of an insulating layer (yellow), and modulates the tunneling current through the spin-polarized (SP) tip
under micro-wave voltage drive $V_{\mathrm{rf}}$. Upper-right inset: Schematic plot of the low-energy spectrum of a typical probe spin with $S=2$ (e.g., Fe on MgO) from the effective spin Hamiltonian Eq.~(\ref{EffectSpinHam}), see also Fig.~\ref{FIG7 FeEnSpec} in Appendix \ref{Sec Appendix C}. The transition frequency $f_{0}$ between the low-energy states $|0\rangle$ and $|1\rangle$, modified by the dynamical spin fluctuations of the Kondo system, is of primary interest in this study. 
}\label{FIG 1 Schematics}
\end{figure}

The setup we consider here is based on a combination of electron spin resonance and scanning tunneling microscopy (ESR-STM, see Fig.~\ref{FIG 1 Schematics}) \cite{2015_Science_Baumann,2017_Nat_Nano_Choi,2018_Sci_Adv_Willke}. 
The key idea is to detect the small level shifts in the probe spin due to the
spin fluctuations of the Kondo impurity. In essence, the level shifts are similar to the Lamb shift of atomic levels due to the quantum fluctuations of the electromagnetic
field \cite{1947_PR_Lamb,1947_PR_Bethe}. In fact, the charge compressibility
measurement \cite{2017_Nature_Desjardins} was along the same principle, by
detecting the shift of the cavity resonance frequency due to the \emph{charge}
fluctuations on the quantum dot. In that case, the bosonic nature
(i.e. linear system) of the cavity photons allows one to estimate the shift by
means of the random phase approximation or similar methods. 

In our case, the probe spin is a highly non-linear system and the theoretical analysis is more complicated. We express and estimate the level shifts in
terms of the dynamic spin susceptibility based on the open quantum system
approach, by regarding the probe spin as the ``system'' and the Kondo impurity
spin as the fluctuating ``bath''. Testing various common adatoms (as the
probe spin) on surfaces, we illustrate that the level shifts are within the resolution of ESR-STM. We find that transverse
magnetic anisotropy (in addition to the common axial one) of the probe spin
enhances the level shifts.


This article is organized as follows: In Sec.~\ref{Sec II} we present the model of the Kondo impurity and discuss how it is coupled to, and detected through, the ESR-STM-based probe spin.
In Sec.~\ref{SecIII} we develop the general theory underlying this work. We derive the shifts in the energy levels of the probe spin, and connect them to the dynamical spin susceptibility of the Kondo impurity.
Then, those general discussions are applied in Sec.~\ref{Sec IV} to specific
examples from various common impurities appeared in the literature.
Section~\ref{conclusion} concludes the article with some additional remarks.
To maintain the major scope of the work clearer, we defer some technical details to Appendices \ref{Sec ESRSTM theory}, \ref{Sec ME derivation}, and \ref{Sec Appendix C}.

\section{Setup and model\label{Sec II}}

We propose to utilize the ESR-STM techniques to perform an accurate spectroscopic analysis of spin fluctuations in the Kondo effect.
Our ESR-STM-based setup is schematically shown in Fig.~\ref{FIG 1 Schematics} and consists of a Kondo impurity magnetically coupled to a probe spin.
It is governed by the Hamiltonian
\begin{equation}
\hat{H}=\hat{H}_{\mathrm{Kondo}}+\hat{H}_{\mathrm{probe}}+\hat{H}_{\mathrm{exchange}},\label{FullHam}
\end{equation} 
where the three different terms refer to the system, probe, and system-probe interaction Hamiltonians, respectively.
$\hat{H}_{\mathrm{exchange}}$ is typically small, and the Kondo impurity can be treated as providing an effective environment to the probe spin.
The fluctuating environment, i.e., the Kondo impurity, 
induces small shifts to the energy levels of the probe spin.
Such shifts will then be reflected
in
the position of ESR lines detected by the STM tip. The aim of the setup is therefore to indirectly probe the spin fluctuations of the Kondo system through the displaced ESR resonance line of the probe spin.
We will provide below the theoretical framework to compute such effects.

\subsection{Probe impurity and ESR-STM detection}

The spin of the probe impurity (hereafter simply called the \emph{probe spin}) is coupled to the metal substrate through an insulating layer as shown in Fig.~\ref{FIG 1 Schematics}
(while the metallic substrate is necessary for the STM current, the insulating layer prevents the probe spin from being directly coupled to the metallic substrate, which would screen the probe spin).
It is addressed with an STM device.
To a good approximation, several relevant magnetic impurities can be described by the effective spin Hamiltonian \citep{1950_Proc_Phys_Soc_A_Pryce,1953_Rep_Prog_Phys_Bleaney,1999_BOOK_Boca,2012_BOOK_Abragam,2007_Science_Hirjibehedin,2009_PRL_Otte}:
\begin{multline}
\hat{H}_{\mathrm{probe}}
= D_{s}\hat{S}_{\mathrm{p}}^{z}\hat{S}_{\mathrm{p}}^{z}
+E_{s}(\hat{S}_{\mathrm{p}}^{x}\hat{S}_{\mathrm{p}}^{x}
-\hat{S}_{\mathrm{p}}^{y}\hat{S}_{\mathrm{p}}^{y}) \\
+g\mu_{B}\mathbf{B}\cdot\hat{\mathbf{S}}_{\mathrm{p}},
\label{EffectSpinHam}
\end{multline}
where $\hat{S}_{\mathrm{p}}^{\mu}$ is the $\mu$-th component of the probe-spin angular momentum $(\mu=x,y,z)$, $D_{s}$ and $E_{s}$ are the axial and transverse anisotropy parameters,
resepectively,
\textbf{$\mathbf{B}$} is the magnetic field, and $g$ is the $g$-factor of the probe spin.
Typical values of the parameters from the literature are listed in Table~\ref{Table effective parameters}. Physically, the $D_{s}$ and $E_{s}$ terms arise as second-order corrections from the spin-orbit interaction, in the presence of crystal-field anisotropy induced by the insulating substrate \citep{2015_PRL_Baumann,2015_Science_Baumann}.

\begin{table}
\caption{The parameters in the effective spin Hamiltonian for some reported probe spins.
$D_{s}$ and $E_{s}$ (in unit of meV) are the axial and transverse magnetic anisotropy
parameters in Eq. (\ref{EffectSpinHam}). $g$ is the electron $g$-factor,
$S$ is the collective spin angular momentum. Base A/B indicates the
material for the insulating layer (A) and the metallic STM substrate (B), while
the binding site and local symmetry group (Sym.) of the probe spin are also specified.
 \label{Table effective parameters}}
\centering{}%
\begin{tabular}{cccccccc}
 &  &  &  &  &  &  & \tabularnewline
\hline 
\hline 
Type & $S$ & Base & Site & Sym. & $D_{s}$ & $E_{s}$ & $g$\tabularnewline
\hline 
Mn \citep{2007_Science_Hirjibehedin} & 5/2 & $\mathrm{Cu_{2}N}$/Cu & Cu & $D_{2}$ & -0.039 & 0.007 & 1.90\tabularnewline
Fe \citep{2007_Science_Hirjibehedin} & 2 & $\mathrm{Cu_{2}N}$/Cu & Cu & $D_{2}$ & -1.55 & 0.31 & 2.11\tabularnewline
Ce \citep{2020_PRL_Ternes} & 3/2 & $\mathrm{Cu_{2}N}$/Cu & Cu & $D_{2}$ & -1.3 & 0.18 & \tabularnewline
\hline 
Fe \citep{2015_Science_Baumann,2015_PRL_Baumann,2016_Nat_Phys_Paul} & 2 & MgO/Ag & O & $C_{4v}$ & -4.7 & 0 & 2.6 \tabularnewline
Co \citep{2008_Nat_Phys_Otte,2009_JPhysCondMat_Ternes,2009_PRL_Otte} & 3/2 & $\mathrm{Cu_{2}N}$/Cu & Cu &  & 2.75 & 0 & 2.19\tabularnewline
Co \citep{2014_Science_Rau} & 3/2 & MgO/Ag & O & $C_{4v}$ & 27.5 &  & \tabularnewline
\hline 
Ti \citep{2017_PRL_Yang,2019_PRL_Yang} & 1/2\footnote{The Ti probe spin in Refs. \onlinecite{2017_PRL_Yang,2019_PRL_Yang} has spin $1/2$ because of a hydrogen atom attached to the Ti adatom. It has been pointed out that for clean Ti on MgO one would have $S=1$ \citep{2017_PRL_Yang}.} & MgO/Ag &  &  & - & - & 1.98\tabularnewline
\hline 
\hline 
 &  &  &  &  &  &  & \tabularnewline
\end{tabular}
\end{table}

The spin-dependent level structure of the probe spin is detected accurately by the ESR-STM measurement with a spin-polarized tip.
Note that the same setup can actively manipulate/detect the spin state of the probe spin as well.
All these are achieved
by feeding a micro-wave drive $V_{\mathrm{rf}}$ of frequency $f$ on the bias voltage of the spin-polarized tip, which induces a time-dependent change to the local electrostatic environment of the probe spin. The drive induces spin transitions via the spin-orbit coupling. We will be interested below in the transition between the two lowest levels of the probe spin, which we denote as $|0\rangle$ (the ground state) and $|1\rangle$ (the first excited state). Since typically $E_s \ll D_s$ (see Table~\ref{Table effective parameters}), at small magnetic field these two states form a doublet. When $D_s<0$:
\begin{equation}
|0\rangle,|1\rangle \simeq |S,\pm S \rangle,
\end{equation}
while for $D_s>0$ and $S$ is half-integer (like in the case of Co) the low-energy states are $|0\rangle,|1\rangle \simeq |S, \pm 1/2 \rangle$. Because the probe spin states involved by the ESR drive differ in their magnetic moments, the spin transition will cause a change in tunneling current $I(f)$
through the probe spin from the spin-polarized tip to the substrate.
Technical details of the tunneling current $I(f)$ are explained in
Appendix~\ref{app:tunnel_current},
following previous literature \cite{2015_Science_Baumann,2009_PRL_Fernandez,2010_PRB_Delgado}.
The signal is approximately given by:
\begin{equation}
\label{If}
I(f) \simeq I_{0}+I_{p}\left[ \frac{\Omega^{2}T_{1}/T_{2}}{4\pi^2(f-f_{0})^{2}+\Omega^{2}T_{1}/T_{2}+1/T_{2}^{2}} \right],
\end{equation}
where $\Omega$ is the Rabi frequency of the drive and $f_{0}:=(E_{1}-E_{0})/2\pi\hbar$ is the ESR transition frequency. $1/T_{1}$ and $1/T_{2}$ are, respectively, the relaxation and dephasing rates of the probe spin.
The explicit expressions for the background ($I_{0}$) and saturation ($I_{p}$) currents are described in the Appendix~\ref{app:tunnel_current} [see Eqs. (\ref{I0Ip})], and include both contributions from elastic and inelastic tunneling \citep{2009_PRL_Fernandez,2010_PRB_Delgado}.


As we explain in detail in Appendix~\ref{app:tunnel_current}, the tunneling current Eq.~(\ref{If}) is derived under the assumption of a small bias voltage between the spin-polarized tip and the substrate, i.e., the bias is much smaller then $D_{s}$. Thus, only the lowest states $|0\rangle,|1\rangle$ of the probe spin need to be considered. Under this two-level approximation, it is known that the change in tunneling current $I(f)$ induced by the drive is simply proportional to the steady-state population $P_{1}(f)$ of the upper level \citep{2015_Science_Baumann}. When the incoherent excitation rate $\Gamma_{0\rightarrow1}$ from the ground state $|0\rangle$ to the excited state $|1\rangle$ is negligible, a simplified expression for $P_1(f)$ can be obtained by solving the Bloch equations \citep{2015_Science_Baumann}, and is given by the square parenthesis of Eq.~(\ref{If}). As a result, the ESR signal $I(f)$ has a Lorentzian lineshape centered around the $0$-$1$ transition frequency, which allows to map out the ESR spectrum by sweeping the drive frequency $f$.

\subsection{Kondo impurity}

The Kondo impurity is shown in the right dashed box of Fig.~\ref{FIG 1 Schematics}. Unlike the probe spin
(coupled to an insulating host),
it is strongly coupled to the metallic substrate, which justifies neglecting anisotropy effects
(an anisotropy in the Kondo coupling is irrelevant in the renormalization group sense as the strong-coupling fixed point is isotropic).
One could describe it with a standard version of the Anderson model \citep{1993_BOOK_Hewson}:
\begin{multline}
\label{H_anderson}
\hat{H}_{\mathrm{Kondo}}= \left(\sum_{\sigma}\epsilon_{\sigma}\hat{n}_{\mathrm{d},\sigma}+U\hat{n}_{\mathrm{d},\uparrow}\hat{n}_{\mathrm{d},\downarrow}\right)+\sum_{\mathbf{k},\sigma}\varepsilon_{\mathbf{k}}\hat{c}_{\mathbf{k},\sigma}^{\dagger}\hat{c}_{\mathbf{k},\sigma} \\{} +\sum_{\mathbf{k},\sigma}V_{\mathbf{k}}\left(\hat{d}_{\sigma}^{\dagger}\hat{c}_{\mathbf{k},\sigma}+\hat{c}_{\mathbf{k},\sigma}^{\dagger}\hat{d}_{\sigma}\right)
\end{multline}
where the first term (in the brackets) refers to the Kondo impurity, the second to the substrate, and the third (second line) to the electron tunneling between the Kondo impurity and the substrate. In Eq.~(\ref{H_anderson}), $\hat{d}_{\sigma}$ and $\hat{c}_{\mathbf{k},\sigma}$ are annihilation operators for the impurity and substrate, respectively, and $\hat{n}_{\mathrm{d},\sigma}=\hat{d}_{\sigma}^{\dagger}\hat{d}_{\sigma}$ gives the occupation
of the Kondo impurity level
by an electron with spin $\sigma = \uparrow,\downarrow$.
Here we assume that the Kondo impurity has spin $S_\mathrm{K}=1/2$.
For larger spins, the Kondo effects are even richer. However, the dynamic spin fluctuations due to such exotic Kondo effects can also be detected in the same method, and the resulting frequency shift can be estimated in an essentially similar method described below.

A crucial quantity characterizing magnetic fluctuations of the Kondo impurity
and hence affecting the ESR-STM signal,
is the dynamical spin susceptibility:\cite{endnote:3}
\begin{equation}
\label{DynSusp}
i\hbar\chi_{\mu}(\omega) = -(g\mu_B)^2\int_{0}^{\infty}dt\,
e^{i\omega t}
\left\langle[\hat{S}_{\mathrm{K}}^{\mu}(t),\hat{S}_{\mathrm{K}}^{\mu}(0)]
\right\rangle \,,
\end{equation} 
where the spin angular momentum operator of the Kondo impurity is given by
\begin{math}
\hat{S}_{\mathrm{K}}^{\mu}
:= \frac{1}{2}\sum_{s,s'}\hat{d}_{s}^{\dagger}\sigma_{s,s'}^{\mu}\hat{d}_{s'},
\end{math}
with $\sigma^{\mu}$ the $\mu$-th Pauli matrix, and $\hat{S}_{\mathrm{K}}^{\mu}(t)=e^{i\hat{H}_{\mathrm{Kondo}}t/\hbar}\hat{S}_{\mathrm{K}}^{\mu}e^{-i\hat{H}_{\mathrm{Kondo}}t/\hbar}$.
While $\chi_{\mu}(\omega)$ is
highly non-trivial to compute theoretically and difficult to access experimentally,
the Bethe Ansatz method
provides one relevant result \cite{1995_inBook_Andrei}
\begin{equation}
\label{chi_Lorenzian}
\chi_\mu(0) = \frac{(g\mu_{B})^{2}}{4\pi k_B T_{K}},
\end{equation}
where $g$ is the $g$-factor, $\mu_B$ is the Bohr magneton, and $T_K$ is the Kondo temperature \cite{endnote:2}.
The full dynamic spin susceptibility could be obtained, e.g., through the numerical renormalization group (NRG) method \citep{2015_PRB_Fang}.
A qualitative feature of $\chi_\mu(\omega)$ is the presence of a peak at $\omega=0$ of width $k_B T_{K}/\hbar$, which develops when the temperature of the Kondo system is lowered to $T \leq T_{K}$. Therefore, for qualitative estimations at low temperatures and $|\omega|\lesssim k_B T_{K}/\hbar$, one may resort to the following approximate form (setting $\mu_B$ to unity):
\begin{equation}
\label{LorentzChi}
\chi_{\mu}(\omega) \approx \chi_\mu(0)
\frac{i k_BT_K}{\hbar\omega+ i k_B T_{K}},
\end{equation}
where $\mathrm{Re}\chi_{\mu}(\omega)$ gives a Lorentzian peak and $\mathrm{Im}\chi_{\mu}(\omega)$ follows from the
the Kramers-Kronig relations. As we will see, it is $\mathrm{Im}\chi_{\mu}(\omega) $ which determines the shift in ESR frequencies of the probe spin. 

Importantly, the sole characteristic energy scale of the Kondo impurity, i.e., the Kondo temperature $T_{K}$, is large in our problem. For example, it is known that the Kondo temperatures of a Co impurity can reach nearly $100~\mathrm{K} \sim 9~\mathrm{meV}$ when deposited without an insulating layer on Ag(111) \citep{2004_PRL_Wahl}. Among Kondo systems with an insulating layer, despite $T_{K}$ may significantly degrade as a result of partial decoupling to the metallic surface, moderate values of $\sim 15~\mathrm{K}$ have been reported for Ce clusters deposited on a monolayer of $\mathrm{Cu}_{2}\mathrm{N}$ on top of Cu(100) \citep{2020_PRL_Ternes}.

One needs to apply an external magnetic field for
the ESR-STM technique, but we work here in a regime where the Zeeman splitting of the Kondo spin is much smaller than $k_BT_K$. As a reference, typical values of the magnetic field $B$ applied in ESR-STM experiments are around $5 $~T \cite{2015_Science_Baumann,2017_Nat_Nano_Choi}, giving $g\mu_B B/k_B \sim 7$~K.
If for a given system the Kondo temperature is not sufficiently large, and working at smaller values of $B$ is not an option, another interesting possibility is to modify the local magnetic environment by introducing additional magnetic impurities.
It was demonstrated that the effect of nearby magnetic impurities can offset the external magnetic field acting on the Kondo spin, and a single Kondo peak can be observed at a finite value of $B$ \cite{2009_PRL_Otte}.

\subsection{Interaction between impurity spins}

To probe the spin fluctuations of the Kondo impurity,
a sizable coupling is necessary
between the probe impurity (spin) and the Kondo impurity, which could be
mediated through the substrate or come from the direct dipole-dipole
coupling. The corresponding Hamiltonian is written as
\begin{equation}
\hat{H}_{\mathrm{exchange}}=\sum_{\mu\in\{x,y,z\}}J_{\mu}\hat{S}_{\mathrm{p}}^{\mu}\hat{S}_{\mathrm{K}}^{\mu},\label{ProbeKondoIntHam}
\end{equation}
which describes either isotropic ($J_{\mu}=J$) or XXZ ($J_{z}\ne J_{x}=J_{y}$ ) interactions. Depending on the system and underlying mechanism,
previously reported coupling strengths vary in a wide range.
Closely spaced impurities can exhibit strong magnetic interactions, dominated by direct exchange, which are able to destroy or strongly modify the Kondo effect \cite{2007_PRL_Wahl,2009_PRL_Otte}. For our purpose we will avoid this regime, by requiring $J_\mu \ll k_B T_K$. Dipolar couplings between two adatoms detected by ESR-STM spectroscopy lie in a suitable range, with reported ESR frequency shifts of $0.02 -3$ GHz \citep{2017_Nat_Nano_Choi} (translating to $J_\mu \sim 0.1-10~\mu$eV). The ESR shifts induced by such dipolar interaction are thus negligibly smaller than the bare transition frequency, $f_0 \simeq 23$~GHz in Ref.~\cite{2017_Nat_Nano_Choi}.

This parameter regime allows us to treat the Kondo impurity as an additional bath for the probe spin. Since a typical Kondo temperature corresponds to spin-flips on a time-scale of $\mathrm{THz}$, while typical ESR transition frequencies are a few tens of $\mathrm{GHz}$ \cite{2015_Science_Baumann,2017_Nat_Nano_Choi,2017_PRL_Yang}, the separation of time-scales implies that the probe spin cannot resolve individual spin flips of the Kondo impurity, but is only influenced by the average effect of spin-fluctuations. The scenario is similar to the open-system dynamics in the presence of a Markovian bath, where only the fast varying bath correlation functions enter the equation of motion \cite{2002_BOOK_Breuer}. An explicit application of this approach to our system is pursued in the next section.

It is optimal to put the Kondo impurity directly on top of the metallic
substrate for stronger coupling \cite{endnote:1}
whereas the probe spin is better absorbed on a thin insulating layer for
sharper ESR resonance lines
(to prevent it from being screened by the Kondo effect).
On the other hand, the Kondo and probe spins should be close enough to have a substantial exchange coupling $J_{\mu}$.
One way to achieve these goals is to employ insulating nano-islands, which have been realized with $\mathrm{Al}_{2}\mathrm{O}_{3}$, $\mathrm{Cu_{2}N}$ or $\mathrm{MgO}$ \cite{2004_Science_Heinrich,2007_Science_Hirjibehedin,2015_PRL_Baumann}.
Positioning an adatom at the edge of an insulating nano-island, where a significant Kondo effect appears, is another way and has also been realized in experiments \cite{2013_Nat_Nano_Oberg}.

\section{Effective environment of the probe spin}
\label{SecIII}

The time-scale separation and weak coupling between the probe spin and Kondo impurity justify treating the Kondo impurity as an effective bath. Following standard procedures \cite{2002_BOOK_Breuer}, the master equation for the reduced density operator $\hat{\rho}_{\mathrm{p}}(t)\equiv\mathrm{Tr}_{\mathrm{Kondo}}\{\hat{\rho}(t)\}$ of the probe spin can be derived as
\begin{multline}
\frac{d}{dt}\hat{\rho}_{\mathrm{p}}^{(\mathrm{I})}(t) =
- \frac{1}{\hbar^2}\int_{0}^{\infty}d\tau\,
\operatorname*{Tr}_{\mathrm{Kondo}}
\left\{\left[\hat{H}_{\mathrm{exchange}}^{(\mathrm{I})}(t),\right.\right. \\
\left.\left.
\left[\hat{H}_{\mathrm{exchange}}^{(\mathrm{I})}(t-\tau),\hat{\rho}_{\mathrm{p}}^{(\mathrm{I})}(t)\otimes\hat{\rho}_{\mathrm{Kondo}}\right]\right]\right\},
\label{Rk}
\end{multline}
where the superscript $\mathrm{I}$ indicates that the operators are defined under the interaction picture with respect to $\hat{H}_{\mathrm{probe}}+\hat{H}_{\mathrm{Kondo}}$, and
the thermal equilibrium $\hat{\rho}_{\mathrm{Kondo}}\propto\exp[-\beta\hat{H}_{\mathrm{Kondo}}]$
is assumed for the Kondo impurity.
The explicit evaluation of Eq.~(\ref{Rk}) is discussed in Appendix \ref{Sec ME derivation}, with the final form including two effects:
\begin{equation}
\label{master_eq}
\frac{d}{dt}\hat{\rho}_{\mathrm{p}}^{(\mathrm{I})}(t) =
\frac{1}{i\hbar}
[\hat{H}_{\mathrm{Lamb}},\hat{\rho}_{\mathrm{p}}^{(\mathrm{I})}(t)]+\mathcal{R}[\hat{\rho}_{\mathrm{p}}^{(\mathrm{I})}(t)].
\end{equation}
The dissipative term $\mathcal{R}[\hat{\rho}_{\mathrm{p}}^{(\mathrm{I})}]$ is given by Eq.~(\ref{LindbladRk}) and describes the
relaxation and dephasing induced by the Kondo system.
It is expressed and estimated more explicilty in Appendix~\ref{Sec ME derivation}.
In principle, the dissipative term leads to a finite resonance line width in the ESR-STM experiment. However, in reality the ESR line width is determined by the instrumental factors such as the strong driving RF field and the tip-sample vibrations \cite{2015_Science_Baumann}. For this reason, we do not further discuss the dissipative term in this work and focus on the coherent (Lamb shift) term.
The shift in the probe spin energy levels $E_m$ is:
\begin{equation}
\hat{H}_{\mathrm{Lamb}}=\sum_{m} \delta E_{m} |m\rangle\langle m|,\label{HLamb}
\end{equation}
where $|m\rangle$ is the eigenstate of $\hat{H}_{\mathrm{probe}}$ with energy $E_m$.
By explicit evaluation (see Appendix~ \ref{Sec ME derivation}), we obtain:
\begin{equation}
\delta E_{m} = -\sum_{\mu}\frac{J_{\mu}^{2}}{\pi k_B T_K}\sum_{n}|S_{nm}^{\mu}|^{2}W_{\mu}(E_{nm}),\label{EnergyShifts}
\end{equation}
where
\begin{math}
E_{nm} := E_{n}-E_{m}
\end{math}
and
\begin{math}
S_{mn}^{\mu}=\langle m|\hat{S}_{\mathrm{p}}^{\mu}|n\rangle
\end{math}
is the spin transition matrix element of the probe spin. The energy weighting factor $W_{\mu}(E)$ is given by:
\begin{align}
W_{\mu}(E) =
\mathrm{Pr}\int_{-\infty}^{\infty} \frac{d\omega}{\pi}
\frac{f_{B}(\hbar\omega/k_B T)\mathrm{Im}\left[ \chi_{\mu}(\omega)/ \chi_\mu(0)\right]}{E/\hbar -\omega} , \label{Wmu}
\end{align}
where $f_{B}(z)=(e^{z}-1)^{-1}$ and the symbol $\mathrm{Pr}$ means the Cauchy principle value. 
Note that, in the regime of our interest, the ratio $\chi_\mu(\omega)/\chi_{\mu}(0)$ is a universal function of $\hbar\omega/k_B T_K$.
From these expressions we see that the energy shifts $\delta{E}_m$ depend crucially on two factors: (i) the spin transition matrix elements $S_{mn}^{\mu}$ and (ii) the dynamical susceptibility $\chi_{\mu}(\omega)$,
contained in the weighting factor $W_{\mu}$. 

\begin{figure}
\begin{centering}
\includegraphics[width=8.6cm]{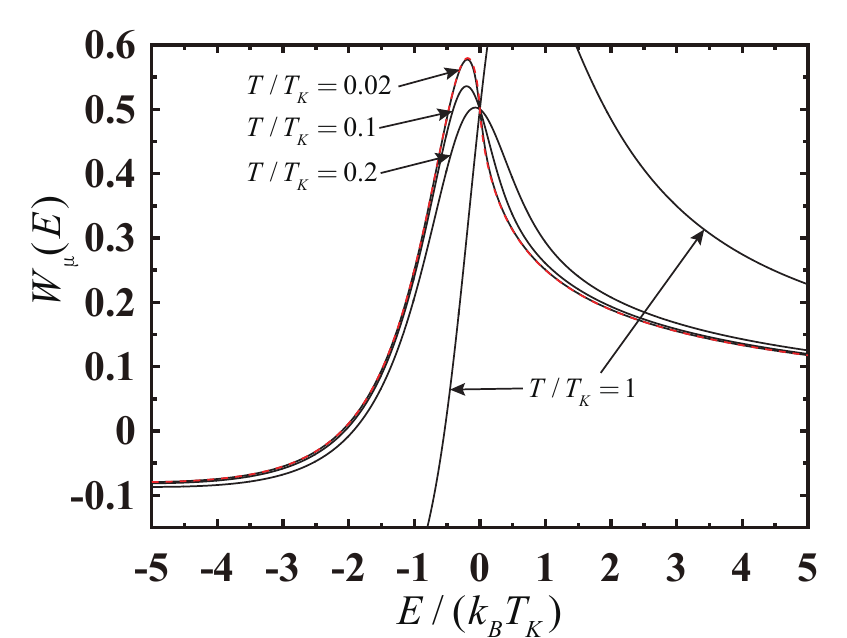}
\end{centering}
\caption{Energy weighting function $W_{\mu}(E)$, defined in Eq.~(\ref{Wmu}), and evaluated under the Lorentzian approximation. The black curves are from a numerical evaluation of Eq.~(\ref{W_lorenzian}) while the red dashed curve is the zero temperature analytical result Eq.~(\ref{WmuZeroTemp}).\label{FIG3 Wmu}}
\end{figure}

For concreteness, we consider the Lorentzian approximation of $\chi_{\mu}(\omega)$, Eq.~(\ref{LorentzChi}), giving:
\begin{align}\label{W_lorenzian}
W_{\mu}(E) =\mathrm{Pr} \int_{-\infty}^{\infty} \frac{dx}{\pi} 
 \left[\frac{x f_{B}(x/\tau)}{1+x^{2}}\right] \frac{1}{\varepsilon-x},
\end{align}
where $\tau=\frac{T}{T_{K}}$ and $\varepsilon=\frac{E}{k_B T_{K}}$.
Figure~\ref{FIG3 Wmu} shows the general behaviors of the energy weighting factor $W_\mu(E)$ as a function of energy for different temperatures.
In the zero temperature limit we obtain:
\begin{equation}
W_{\mu}(E) =\frac{\pi/2+\varepsilon \ln|\varepsilon|}{\pi(1+\varepsilon^{2})},\quad (\tau \rightarrow0)\label{WmuZeroTemp}
\end{equation}
which works well if $T/T_{K}\lesssim10^{-2}$ (cf. Fig.~\ref{FIG3 Wmu}). Another interesting limit is for $\varepsilon \to 0$, giving $W_{\mu}(E=0) =1/2$. The $E=0$ result is independent of $\tau$, also seen in Fig.~\ref{FIG3 Wmu}.

Finally, the shift in spin transition frequency $f_{0}:=(E_{1}-E_{0})/2\pi\hbar$, which can be detected using an ESR-STM device, is obtained as:
\begin{align}
\label{ESRShifts}
\delta f_{0}=\frac{J_{\mu}^{2}/(2\pi\hbar)}{\pi k_B T_K}
\sum_{\mu,n}\left[
|S_{n0}^{\mu}|^{2}W_{\mu}(E_{n0}) -
|S_{n1}^{\mu}|^{2}W_{\mu}(E_{n1})
\right].
\end{align}
To analyze the behavior of $\delta f_{0}$, we will carry out in the next section the numerical evaluations of these expressions
for typical probe spins.

\section{ESR transition frequencies}\label{Sec IV}

\begin{figure}
\begin{centering}
\includegraphics[width=8.6cm]{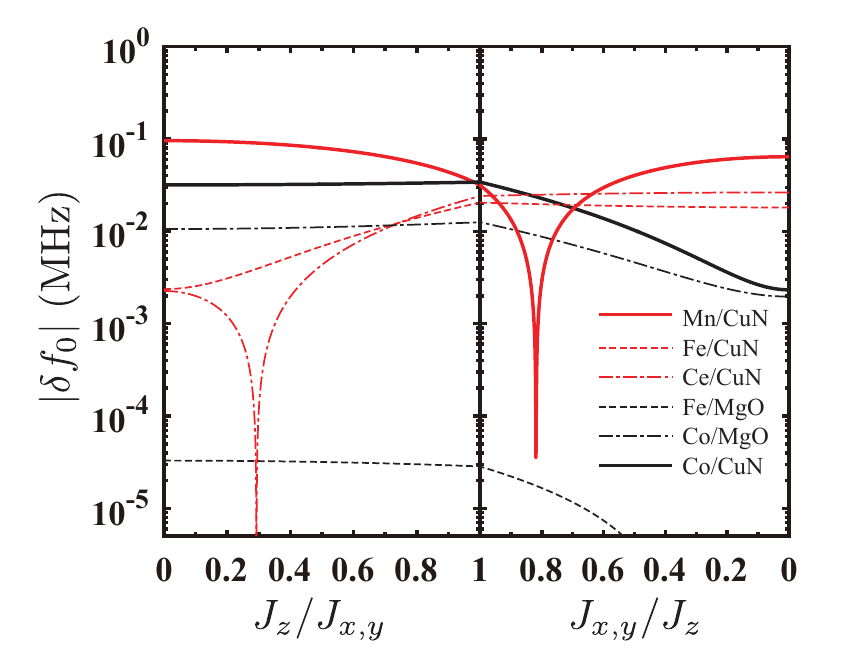}
\par\end{centering}
\caption{Frequency shifts $\delta f_{0}$ as a function of exchange coupling
anisotropy for probe spins listed in Table~\ref{Table effective parameters} (we use $g=2$ if the g-factor is not specified).
Other parameters: $B_{x}=6~\mathrm{T}$, $B_{z}=0.2~\mathrm{T}$,
$T=0.5~\mathrm{K}$, and $k_B T_{K}=1~\mathrm{meV}$. In the left panel $J_{x,y}=1~\mu\mathrm{eV}$
and in the right panel $J_{z}=1~\mu\mathrm{eV}$.\label{FIG 2 typical adatom shifts}}
\end{figure}

We show in Fig.~\ref{FIG 2 typical adatom shifts} the frequency shifts obtained for the probe spins listed in Table~\ref{Table effective parameters}. The shifts have a sensitive dependence on the type of impurity, insulating layer, and substrate,
as well as the anisotropy of the exchange interaction.
We generally find that Mn on $\mathrm{Cu}_{2}\mathrm{N}$ is the most sensitive probe spin, while Fe on MgO always yields a very small shift and will be considered separately below. With this exception, the range of shift agrees reasonably well with the general scale predicted by Eq.~(\ref{ESRShifts}), where the prefactor evaluates to $\simeq 80$~kHz with $J_\mu \simeq 1~\mu$eV and $k_B T_K =1$~meV (the parameters assumed in Fig.~\ref{FIG 2 typical adatom shifts}). In particular, in the isotropic case ($J_z=J_{x,y}$) all the shifts are in the range of $10-25$ kHz, with the variation between different probe spins due to the spin operator matrix elements and weighting factors appearing in Eq.~(\ref{ESRShifts}).

While it is immediate to understand how the overall frequency scale is influenced by a change of $J_\mu$ and $T_K$, the effects related to $|S^\mu_{nm}|^2 W_\mu(E_{nm})$ are much less transparent. For example, sharp dips are induced at certain values of the $J_{z}/J_{x,y}$ ratio, due to a sign change of $\delta f_{0}$. The precise condition for these vanishing shifts, i.e., $\sum_{n}|S_{n1}^{\mu}|^{2}W_{\mu}(E_{n1})$ equals to $\sum_{n}|S_{n0}^{\mu}|^{2}W_{\mu}(E_{n0})$, depends in a nontrivial way on the magnetic field, exchange coupling, as well as temperature. It is also interesting to notice that some probe spins (in particular, Mn and Co) prefer an in-plane exchange coupling $\propto \hat{S}_{\mathrm{p}}^{x}\hat{S}_{\mathrm{K}}^{x}+\hat{S}_{\mathrm{p}}^{y}\hat{S}_{\mathrm{K}}^{y}$ , which facilitates a larger $\delta f_{0}$, while in other cases (Fe and Ce on Cu${}_2$N) the limit of an Ising interaction with $J_{x,y}=0$ is most favorable.

We now return on the special case of Fe on MgO which, despite being one of the commonly employed probe spins in ESR-STM devices, shows the smallest energy shifts among those listed in Fig. \ref{FIG 2 typical adatom shifts}.
From the point of view of the spin Hamiltonian Eq.~(\ref{EffectSpinHam}), the smallness of $\delta f_0$ is attributed to the large spin of the impurity, $S=2$, and the high symmetry of the binding site, which does not allow the transverse anisotropy term ($E_s=0$).
To confirm it, in Appendix~\ref{Sec Appendix C}
we considering a more microscopic Hamiltonian, commonly used in modeling ESR-STM studies \cite{2015_Science_Baumann}.
Under the conditions of large $S$ and uniaxial anisotropy ($E_s=0$),
the unperturbed Hamiltonian $D_{s}\hat{S}_{\mathrm{p}}^{z}\hat{S}_{\mathrm{p}}^{z}$ leads to a low-energy doublet $|S,\pm 2\rangle$, for which the direct matrix elements of the spin operators vanish, i.e., $|S_{01}^\mu|^2=0$ in Eq.~(\ref{ESRShifts}). While the perturbation
\begin{math}
\delta\hat{V}=-g\mu_{B}B_{x}\hat{S}_{\mathrm{p}}^{x}
\end{math}
induced by a transverse field allows tunneling between the two levels, this can only occur through higher order processes. The off-diagonal matrix elements $|\delta V_{S,-S}|$ between
$|S,\pm S\rangle$ had been calculated in the literature, which in the large $S$ limit reads \citep{1991_J_Phys_A_Garanin,1997_PRB_Garanin}
\begin{equation}
|\delta V_{S,-S}|\approx\frac{2D_{s}S^{3/2}}{\pi^{1/2}}\left(\frac{eg\mu_{B}B_{x}}{4SD_{s}}\right)^{2S},
\end{equation}
and decreases rapidly with $S$, thus leading to small values of $|S_{01}^\mu|^2$. While this matrix element is small, $|S_{0n}^\mu|^2$ and $|S_{1n}^\mu|^2$ can become sizable when considering the contributions to Eq.~(\ref{ESRShifts}) from excited states, i.e., $n \neq 0,1$. However, these contributions are suppressed by the unfavorable dependence of the weighting factor by the large transition frequencies $E_{0n}, E_{1n}\sim D_s$.

In passing, we note how the other impurity spins listed in Table~\ref{Table effective parameters} avoid the aforementioned problems of Fe on MgO.
For probe spins at binding sites with low symmetry (described by the point group $D_2$), the finite value of $E_s$ breaks the spin selection rule and allows relatively large values of $|S_{01}^\mu|^2$. It mostly results in large shifts for probe spins such as Mn, Fe and Ce (see the first three rows in Table~\ref{Table effective parameters}).
On the other hand, although Co are sitting on the binding sites with a high symmetry (the point group $C_{4v}$), the positive value of $D_s$ leads to a low-energy doublet of the type $|0\rangle,|1\rangle \simeq |3/2, \pm 1/2 \rangle$. In this case, the spin transition matrix elements are sizable and leads to relatively large level shifts.
Finally, the larger levels shift predicted in Mn on Cu${}_2$N are related to the small values of $D_s, E_s \ll k_ B T_K$ (see Table~\ref{Table effective parameters}), allowing for considerable contributions from virtual transitions through some excited states.

\begin{figure}
\begin{centering}
\includegraphics[width=8.6cm]{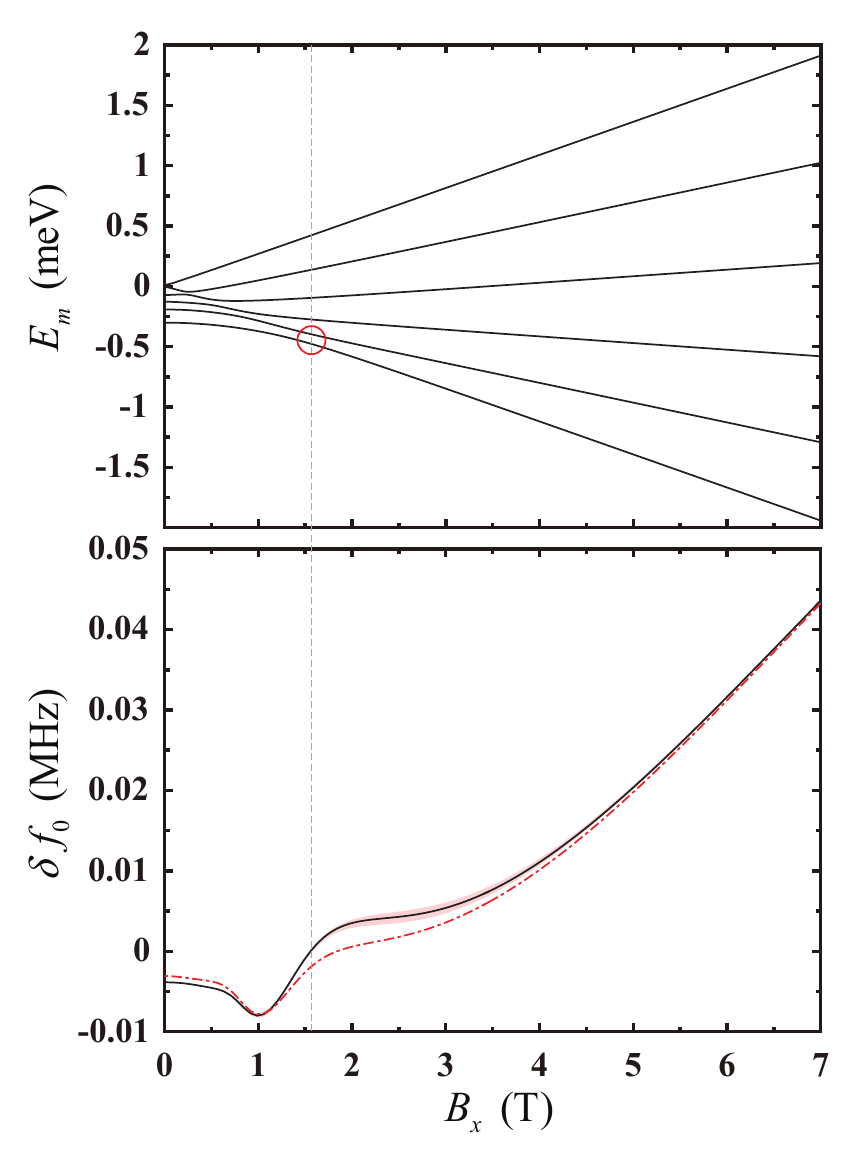}
\par\end{centering}
\caption{Lower panel: Frequency shift $\delta f_{0}$ as a function of magnetic field $B_x$ in
the $x$-direction, with $B_{z}=0.2~\mathrm{T}$. Parameters $D_{s}=-0.039\mathrm{meV}$, $E_{s}=0.007\mathrm{meV}$ and $g=1.90$
correspond to a Mn probe spin. We assumed here an isotropic magnetic interaction, $J_{x}=J_{y}=J_{z}=1~\mathrm{\mu eV}$, and $k_{B}T_{K}=1~\mathrm{meV}$. The solid line is computed for $T=0.5~\mathrm{K}$, with the red shaded region showing the variation
on $\delta f_{0}$ caused by $\Delta T=0.1~\mathrm{K}$ temperature change. The red dash-dotted curve is for $T=0$. Upper panel: Energy spectrum of the probe spin. \label{FIG3 ESRshiftsBx}}
\end{figure}

\begin{figure}
\begin{centering}
\includegraphics[width=8.6cm]{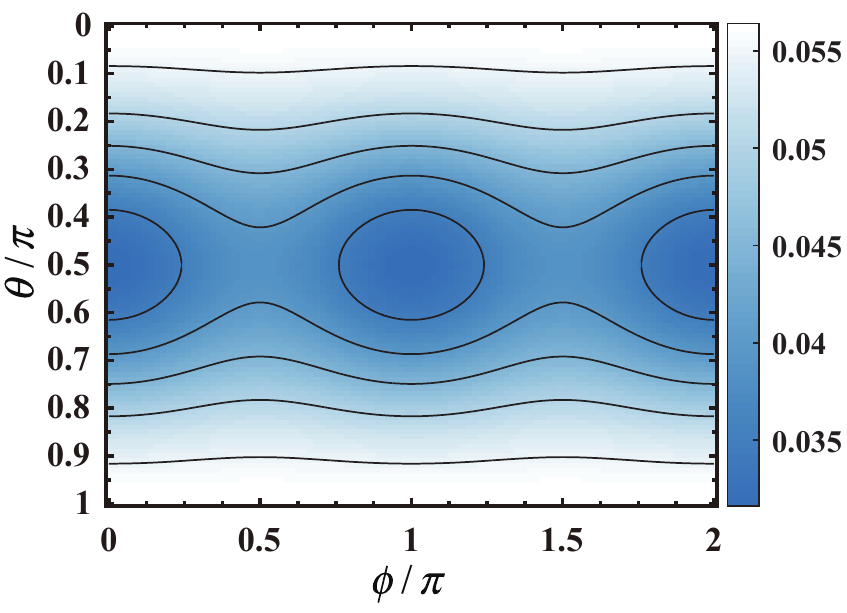}
\par\end{centering}
\caption{Orientation dependence of the frequency shift $|\delta f_{0}|$ for
a magnetic field with a constant magnitude $B=6~\mathrm{T}$. $\theta$
and $\phi$ are the polar and azimuth angles that specify the direction
of $\mathbf{B}$. The color bar is in units of MHz. Except $B_{x,y,z}$,
other parameters are the same as in Fig.~\ref{FIG3 ESRshiftsBx}.\label{FIG4 ESRshiftsOrientations}}
\end{figure}

Before concluding this section,
we examine in more detail the frequency shift in the favorable case of a Mn probe spin on $\mathrm{Cu}_{2}\mathrm{N}$. The lower panel of Fig.~\ref{FIG3 ESRshiftsBx} shows the magnetic field dependence of $\delta f_{0}$, which has a complicated behavior and sign change for $B\leq 3~\mathrm{T}$. A comparison to the low energy spectrum (upper panel) suggests a connection between $\delta f_{0}=0$ and level anti-crossing in the lower-energy doublet.
On the other hand, the behavior of $\delta f_{0}$ at $B\geq 3~\mathrm{T}$ is dominated by the Zeeman term, and shows a nearly quadratic increase of $|\delta f_{0}|$ versus $B_{x}$. Furthermore, since the energy splitting of the probe spin depends on the interplay between the Zeeman term and the magnetic anisotropies, a significant dependence of $\delta f_{0}$ on $\mathbf{B}/B$ may be expected for probe spins with biaxial anisotropy ($E_{s}\ne0$), such as Mn on $\mathrm{Cu}_{2}\mathrm{N}$. This is evident from Fig.~\ref{FIG4 ESRshiftsOrientations}, showing that when the magnetic field is pointing in the $x$-$y$
plane ($\theta=\pi/2$), $|\delta f_{0}|$ can increase by $\sim 20\%$ by tuning the magnetic field direction from the hard axis ($\phi=0,\pi$) to $\phi=\pi/2,3\pi/2$. It is also clear from Fig.~\ref{FIG4 ESRshiftsOrientations} that the largest shift $\delta f_{0}$ is induced by a field $\mathbf{B}$ pointing along the easy axis ($\theta=0,\pi$).

\section{Conclusion}\label{conclusion}

In this work, we have explored a possibility for accessing the spin fluctuations from a Kondo impurity, based on the ESR-STM technique.
The spin fluctuations in the Kondo impurity effectively shift the resonance transition line in the probe spin, which can be detected with the ESR-STM setup \cite{2007_Science_Hirjibehedin,2015_Science_Baumann}.
We estimate theoretically the shifts in the resonance line for various types of probe spins on top of surfaces and express then in terms of the dynamical spin susceptibility.

We find that the expected amount of shifts on the transition frequency $f_0$ are most favorable under the existence of a suitable transverse magnetic anisotropy, or with $|\Delta S^z|=1$ selection rules for the ESR-active states. These findings are similar to the previous studies on Kondo effect with the STM setups where, due to magnetic anisotropy, the conditions for the Kondo resonance from single impurity on metallic surface were shown to be different for impurities with $S>1$ \cite{2008_Nat_Phys_Otte,2009_JPhysCondMat_Ternes}. Through comparison among various existing impurities employed in the ESR-STM studies, it turns out that Mn on $\mathrm{Cu}_2\mathrm{N}$ might be suitable for detecting the spin fluctuations.

Despite of the small frequency shift $\delta f_0$ for Fe on MgO, this type of ESR probe is very promising because of its large perpendicular magnetic anisotropy and the long life-time of spin excited state \cite{2014_Science_Rau,2015_PRL_Baumann,2016_Nat_Phys_Paul}. We expect that significant improvement may be achieved, e.g., through multiple-step transitions involving the excited states, similar methods had already been discussed in the context of single molecular magnets \cite{2001_Nature_Leuenberger,2001_PRB_Park,2008_PRB_Tokman,2006_PRB_Shvetsov,2007_PRB_Wu}.


\appendix

\section{Calculation of $I(f)$\label{Sec ESRSTM theory}} \label{app:tunnel_current} 

We summarize in this Appendix the main features of the tunneling current between the STM tip and the substrate, and discuss how it allows to detect the energy levels of the probe spin. We start by the Hamiltonian of the STM measurement device without an ESR drive:
\begin{equation}
\hat{H}_{\mathrm{STM}}=\hat{H}_\mathrm{probe}+\hat{H}_{\mathrm{s}}+H_\mathrm{t}+\hat{V}_{\mathrm{tun}},\label{HamESRSTM}
\end{equation}
where $\hat{H}_\mathrm{probe}$ is given in Eq.~(\ref{EffectSpinHam}), with eivenvalues and eigenstates $E_n$ and $|n\rangle$, respectively. The other terms of $\hat{H}_{\mathrm{STM}}$ include the free Hamiltonians for the SP-tip and substrate:
\begin{equation}
\hat{H}_{\mathrm{t}}=\sum_{\mathbf{k},\sigma}\varepsilon^\mathrm{T}_{\mathbf{k},\sigma}\hat{t}_{\mathbf{k},\sigma}^{\dagger}\hat{t}_{\mathbf{k},\sigma}, \qquad 
\hat{H}_{\mathrm{s}}=\sum_{\mathbf{q},\sigma}\varepsilon^\mathrm{S}_{\mathbf{q},\sigma}\hat{s}_{\mathbf{q},\sigma}^{\dagger}\hat{s}_{\mathbf{q},\sigma},\label{SubstrateHam}
\end{equation}
as well as their tunnel coupling in the presence of the probe spin \citep{2009_PRL_Fernandez,2010_PRB_Delgado}:
\begin{align}
\hat{V}_{\mathrm{tun}}= & \sum_{\mathbf{k},\mathbf{q}}\sum_{\sigma}\left\{ \left(T_{\mathbf{k},\mathbf{q}}\hat{t}_{\mathbf{k},\sigma}^{\dagger}\hat{s}_{\mathbf{q},\sigma}+\mathrm{h.c.}\right)\right.\nonumber \\
 & +\sum_{\mu}\sum_{\sigma'}\left(T'_{\mathbf{k},\mathbf{q}}\frac{\sigma_{\sigma,\sigma'}^{\mu}}{2}\hat{S}_{\mathrm{p}}^{\mu}\hat{t}_{\mathbf{k},\sigma}^{\dagger}\hat{s}_{\mathbf{q},\sigma'}+\mathrm{h.c.}\right) \\
\simeq & \sum_{\alpha}\sum_{\mathbf{k},\sigma}\sum_{\mathbf{q},\sigma'}\left(T^{(\alpha)}\frac{\sigma_{\sigma,\sigma'}^{\alpha}}{2}\hat{S}_{\mathrm{p}}^{\alpha}\hat{t}_{\mathbf{k},\sigma}^{\dagger}\hat{s}_{\mathbf{q},\sigma'}+\mathrm{h.c.}\right).\label{TunnelHam}
\end{align}
The first (second) line of Eq.~(\ref{TunnelHam}) describes
elastic (inelastic) tunneling, with tunneling strength
given by $T_{\mathbf{k},\mathbf{q}}$ ($T'_{\mathbf{k},\mathbf{q}}$). 
In the third line of Eq.~(\ref{TunnelHam}), the index $\alpha$ sums over $0$, and $\mu=x,y,z$,
 with $\hat{S}_{\mathrm{p}}^{\alpha=0}=\hat{I}_{\mathrm{p}}$
the identity operator and $\sigma^0$ the two-by-two identity matrix. In the spirit of the commonly employed wide band limit approximation, the tunnel couplings $T_{\mathbf{k},\sigma}$
and $T_{\mathbf{q},\sigma'}'$ are assumed to be constant
\citep{1994_PRB_Jauho},
i.e., $T^{(0)}\equiv T_{0}$ and $T^{(\mu)}=T_{0}'$
while $\zeta=T_{0}'/T_{0}$ quantifies the relative strength of inelastic
to elastic tunneling. 

In the weak-tunneling limit, the relaxation rate from state $|m\rangle$ to state $|n\rangle$ of the probe spin could be calculated
using Fermi golden rule expressions \cite{2009_PRL_Fernandez,2010_PRB_Delgado}:
\begin{align}
\hbar\Gamma_{m\rightarrow n}^{\mathrm{tip}\rightarrow\mathrm{sub}}= & \frac{\pi}{2}\sum_{\mathbf{k},\mathbf{k}'}\sum_{\sigma,\sigma'}|\sum_{\alpha}T^{(\alpha)}\sigma_{\sigma',\sigma}^{\alpha}S_{nm}^{\alpha}|^{2}[1-f_{\mathrm{S}}(\varepsilon_{\mathbf{k}',\sigma'}^{\mathrm{S}})]\nonumber \\
 & \times f_{\mathrm{T}}(\varepsilon_{\mathbf{k},\sigma}^{\mathrm{T}})\delta(E_{m}+\varepsilon_{\mathbf{k},\sigma}^{\mathrm{T}}-E_{n}-\varepsilon_{\mathbf{k}',\sigma'}^{\mathrm{S}}),\label{RateST}
\end{align}
and
\begin{align}
\hbar\Gamma_{m\rightarrow n}^{\mathrm{sub}\rightarrow\mathrm{tip}}= & \frac{\pi}{2}\sum_{\mathbf{k},\mathbf{k}'}\sum_{\sigma,\sigma'}|\sum_{\alpha}T^{(\alpha)}\sigma_{\sigma,\sigma'}^{\alpha}S_{nm}^{\alpha}|^{2}[1-f_{\mathrm{T}}(\varepsilon_{\mathbf{k},\sigma}^{\mathrm{T}})]\nonumber \\
 & \times f_{\mathrm{S}}(\varepsilon_{\mathbf{k}',\sigma'}^{\mathrm{S}})\delta(E_{m}+\varepsilon_{\mathbf{k}',\sigma'}^{\mathrm{S}}-E_{n}-\varepsilon_{\mathbf{k},\sigma}^{\mathrm{T}}),\label{RateTS}
\end{align}
where $S_{nm}^{\alpha}=\langle n |\hat S^\alpha_\mathrm{p}| m \rangle$ are matrix elements of the probe spin, $f_{\mathrm{S,T}}(\varepsilon)=[\exp[\beta(\varepsilon-\mu_\mathrm{S,T})]+1]^{-1}$ are the Fermi-Dirac distribution function of the substrate and SP-tip, and we take real tunneling amplitudes. $\Gamma_{m\rightarrow n}^{\mathrm{tip}\rightarrow\mathrm{sub}}$
refers to the transition rate of the probe spin from $|m\rangle$ to $|n\rangle$
due to an electron tunneling from tip to substrate while $\Gamma_{m\rightarrow n}^{\mathrm{sub}\rightarrow\mathrm{tip}}$
describes the same transition due to a reverse tunneling. The total transition rate is
\begin{equation}
\Gamma_{m\rightarrow n}=\Gamma_{m\rightarrow n}^{\mathrm{tip}\rightarrow\mathrm{sub}}+\Gamma_{m\rightarrow n}^{\mathrm{sub}\rightarrow\mathrm{tip}}.\label{total rate}
\end{equation}
Following the standard approach of replacing the summation over $\mathbf{k},\mathbf{k}^\prime$ by integrals over energy, and assuming a constant (spin-resolved) density of states $\rho_{\mathrm{T}\sigma}$ and $\rho_{\mathrm{S}\sigma^\prime}$ for the SP-tip and the substrate respectively, we can recast Eqs.~(\ref{RateST}) and (\ref{RateTS}) to the following form:
\begin{align}
\hbar\Gamma_{m\rightarrow n}^{\mathrm{tip}\rightarrow\mathrm{sub}}=\frac{\pi}{2}\sum_{\alpha,\alpha'}T^{(\alpha)}S_{nm}^{\alpha}M_{\alpha\alpha'}G(E_{mn}-eV)T^{(\alpha')}S_{mn}^{\alpha'}, \nonumber \\ 
\hbar\Gamma_{m\rightarrow n}^{\mathrm{sub}\rightarrow\mathrm{tip}}=\frac{\pi}{2}\sum_{\alpha,\alpha'}T^{(\alpha')}S_{nm}^{\alpha'}M_{\alpha'\alpha}G(E_{mn}+eV)T^{(\alpha)}S_{mn}^{\alpha},\label{RateTS1}
\end{align}
where the matrix $M$ contains products of type $\rho_{\mathrm{T}\sigma}\rho_{\mathrm{S}\sigma^\prime}$ and is defined as follows (basis ordering $\alpha=0,x,y,z$):
\begin{widetext}
\begin{equation}
M=\left[\begin{array}{cccc}
\rho_{\mathrm{S}\uparrow}\rho_{\mathrm{T}\uparrow}+\rho_{\mathrm{S}\downarrow}\rho_{\mathrm{T}\downarrow} & 0 & 0 & \rho_{\mathrm{S}\uparrow}\rho_{\mathrm{T}\uparrow}-\rho_{\mathrm{S}\downarrow}\rho_{\mathrm{T}\downarrow}\\
0 & \rho_{\mathrm{S}\uparrow}\rho_{\mathrm{T}\downarrow}+\rho_{\mathrm{S}\downarrow}\rho_{\mathrm{T}\uparrow} & i\rho_{\mathrm{S}\uparrow}\rho_{\mathrm{T}\downarrow}-i\rho_{\mathrm{S}\downarrow}\rho_{\mathrm{T}\uparrow} & 0\\
0 & -i\rho_{\mathrm{S}\uparrow}\rho_{\mathrm{T}\downarrow}+i\rho_{\mathrm{S}\downarrow}\rho_{\mathrm{T}\uparrow} & \rho_{\mathrm{S}\uparrow}\rho_{\mathrm{T}\downarrow}+\rho_{\mathrm{S}\downarrow}\rho_{\mathrm{T}\uparrow} & 0\\
\rho_{\mathrm{S}\uparrow}\rho_{\mathrm{T}\uparrow}-\rho_{\mathrm{S}\downarrow}\rho_{\mathrm{T}\downarrow} & 0 & 0 & \rho_{\mathrm{S}\uparrow}\rho_{\mathrm{T}\uparrow}+\rho_{\mathrm{S}\downarrow}\rho_{\mathrm{T}\downarrow}
\end{array}\right].\label{Mmat}
\end{equation}
\end{widetext}
In Eq.~(\ref{RateTS1}), the function $G(E) =E /(1-e^{-\beta E})$ is due to the overlapping of
the two Fermi-Dirac distribution functions and $eV=\mu_\mathrm{S}-\mu_\mathrm{T}$ is the applied bias, defined such that electrons tunnel from the SP-tip to the substrate
if $V>0$ ($e$ includes sign).

In the absence of ESR drive, the populations $P_{m}=\langle m|\hat{\rho}_{\mathrm{p}}|m\rangle$ of the probe spin in state $|m\rangle$ can be found by solving the rate equations \citep{2009_PRL_Fernandez,2010_PRB_Delgado}:
\begin{equation}
\dot{P}_{m}(t)=\sum_{n}\left(\Gamma_{n\rightarrow m}P_{n}(t)-\Gamma_{m\rightarrow n}P_{m}(t)\right),\label{RateEq}
\end{equation}
which yield the stationary values $P_{m}(\infty)$. The d.c. current $I_{\mathrm{dc}}$ can be evaluated as:
\begin{equation}
I_{\mathrm{dc}}=-e\sum_{m,n}\left(\Gamma_{m\rightarrow n}^{\mathrm{tip}\rightarrow\mathrm{sub}}-\Gamma_{m\rightarrow n}^{\mathrm{sub}\rightarrow\mathrm{tip}}\right)P_{m}(\infty).\label{Idc}
\end{equation}

When the probe spin is driven by micro-wave field at frequency $f$, one has to generalize the rate equation Eq.~(\ref{RateEq}) to a master equation, which leads to a modification of the steady state populations $P_m(\infty)$ (thus, of the tunneling current). To discuss the effect in a simple and relevant limit, we suppose that the temperature and applied bias are sufficiently small to neglect all the excitation rates, i.e., $\Gamma_{m \to n} \approx 0$ if $n>m$. Then, $P_{0}(\infty)=1$ in the absence of the drive and a finite $P_{1}(\infty)$ is induced by the ESR excitation. Under these conditions, the system can be restricted to the lowest two eigenstates $|0\rangle$ and $|1\rangle$ of $\hat{H}_\mathrm{probe}$, with the tunneling current taking the form:
\begin{align}\label{Ifappendix}
I(f)= I_0 + I_p P_{1}(\infty),
\end{align}
where we used Eq.~(\ref{Idc}) together with $P_{0}(\infty)=1-P_{1}(\infty)$. The background ($I_0$) and saturation ($I_p$) currents are the same of Eq.~(\ref{If}) of the main text, and are given by:
\begin{align}
\label{I0Ip}
& I_0 = -e \left(\Gamma_{0\rightarrow 0}^{\mathrm{tip}\rightarrow\mathrm{sub}}-\Gamma_{0\rightarrow 0}^{\mathrm{sub}\rightarrow\mathrm{tip}}\right),\nonumber \\
& I_p=-I_0-e\sum_{n=0,1}\left(\Gamma_{1\rightarrow n}^{\mathrm{tip}\rightarrow\mathrm{sub}}-\Gamma_{1\rightarrow n}^{\mathrm{sub}\rightarrow\mathrm{tip}}\right),
\end{align}
in terms of the transition rates defined in Eq.~(\ref{RateTS1}).

To compute $P_{1}(\infty)$ under the two-level approximation, we describe the dynamical evolution through the Bloch equation \citep{2015_Science_Baumann}:
\begin{multline}
\frac{d}{dt}\hat{\rho}_{\mathrm{p}} =
\frac{1}{i\hbar}
[\hat{H}_{\mathrm{TLS}}(t),\hat{\rho}_{\mathrm{p}}] \\{}
+ \Gamma_{1\rightarrow0}\mathcal{D}[\hat{L}_{0,1}]\hat{\rho}_{\mathrm{p}}
+ 2\gamma_{\mathrm{dp}}\mathcal{D}[\hat{L}_{1,1}]\hat{\rho}_{\mathrm{p}},
\label{DriveME}
\end{multline}
where
$\hat{L}_{m,n}:=|m\rangle\langle n|$,
$\mathcal{D}[\hat{L}]$ is a superoperator defined by
\begin{math}
\mathcal{D}[\hat{L}]\hat{A}:=
\hat{L}\hat{A}\hat{L}^{\dagger}
- \frac{1}{2}
\{\hat{L}^{\dagger}\hat{L},\hat{A}\}$ for a linear operator $\hat{A}.
\end{math}
The unitary dynamics in Eq.~\eqref{DriveME} is determined by
\begin{equation}
\hat{H}_{\mathrm{TLS}}(t)
= \sum_{n=0,1 }E_{n}|n\rangle\langle n|
+ \frac{\hbar\Omega}{2}
\left(|0\rangle\langle 1|e^{i2\pi ft}+\mathrm{h.c.}\right),
\end{equation}
with Rabi frequency $\Omega$. The second line of Eq.~(\ref{DriveME}) describes the dissipative dynamics with phenomenological pure dephasing rate
$\gamma_{\mathrm{dp}}$.
The stationary solution of Eq.~(\ref{DriveME}) is well-known:
\begin{equation}
\label{P1}
P_1(\infty) = \frac{1}{2}\,
\frac{\Omega^2T_{1}/T_{2}}{4\pi^2(f-f_{0})^{2}+\Omega^2 T_{1}/T_{2}+1/T_{2}^{2}},
\end{equation}
where the longitudinal and transverse relaxation rates are respectively given by
\begin{math}
1/T_1=\Gamma_{1\rightarrow0}
\end{math}
and
\begin{math}
1/T_2=\Gamma_{1\to0}/2+\gamma_\mathrm{dp}.
\end{math}
Finally, substituting Eq.~(\ref{P1}) in Eq.~(\ref{Ifappendix}) leads to Eq.~(\ref{If}) of the main text.

\section{Master equation and $\hat{H}_{\mathrm{Lamb}}$\label{Sec ME derivation}}

Here, we give the detailed derivation for the energy shifts Eq.~(\ref{EnergyShifts}). We start from the master equation Eq. (\ref{Rk}), which is the standard form derived under the Born-Markov approximation, valid in the weak-coupling limit \cite{1990_BOOK_Fick,2002_BOOK_Breuer}. It is worth mentioning that, in writing such master equation, we also assume:
\begin{equation}
\langle\hat{S}_{\mathrm{K}}^{\mu}\rangle_{\mathrm{Kondo}}\equiv\mathrm{Tr}_{\mathrm{Kondo}}\{\hat{S}_{\mathrm{K}}^{\mu}\hat{\rho}_{\mathrm{Kondo}}\}=0.
\end{equation}
This condition of a negligible spin polarization is consistent with the general discussion in the main text, where we required that $k_B T_K$ is much larger than other energy scales (and, in particular, than the Zeeman energy).

To obtain an explicit form of the master equation in Eq.~(\ref{Rk}), we further invoke the rotating wave approximation (RWA) under the eigenstate representation of $\hat{H}_{\mathrm{probe}}$ \citep{2002_BOOK_Breuer}, which gives
\begin{align}
\frac{d}{dt}\hat{\rho}_{\mathrm{p}}^{(\mathrm{I})}(t)=
\frac{1}{i\hbar}\sum_{\mu,m,n} \left[|S_{mn}^{\mu}|^{2}\mathcal{W}_{\mu}(E_{nm})|m\rangle\langle m|,\hat{\rho}_{\mathrm{p}}^{(\mathrm{I})}(t)\right]\nonumber \\
+ \frac{2\pi}{\hbar}
\sum_{\mu}J_{\mu}^{2}\left(\sum_{m\ne n}A_{\mu}^{>}(E_{mn}/\hbar)|S_{mn}^{\mu}|^{2}\mathcal{D}\left[|n\rangle\langle m|\right]\right.\nonumber \\
\left.+A_{\mu}^{>}(0)\mathcal{D}\left[\sum_{m}S_{mm}^{\mu}|m\rangle\langle m|\right]\right)\hat{\rho}_{\mathrm{p}}^{(\mathrm{I})}(t),
\label{LindbladRk}
\end{align}
where $A_{\mu}^{>}(\omega)$ is the diagonal part of the greater spectral function:
\begin{equation}
2\pi\hbar A_{\mu\nu}^{>}(\omega)=\int_{-\infty}^{\infty}dt\, e^{i\omega{t}}
\left\langle\hat{S}_{\mathrm{K}}^{\mu}(t)\hat{S}_{\mathrm{K}}^{\nu}
\right\rangle ,
\end{equation}
and the time dependence in the Kondo impurity operators
\begin{math}
\hat{S}_{\mathrm{K}}^{\mu}(t)
= e^{i\hat{H}_{\mathrm{Kondo}}t/\hbar}
\hat{S}_{\mathrm{K}}^{\mu}
e^{-i\hat{H}_{\mathrm{Kondo}}t/\hbar}
\end{math}
is due to the interaction picture.
Note that the greater spectral function is diagonal, $A_{\mu\nu}^{>}(\omega) = A_{\mu}^{>}(\omega) \delta_{\mu\nu}$, due the spin isotropy of the Anderson Hamiltonian Eq.~(\ref{H_anderson}).

The first line of Eq.~\eqref{LindbladRk} is responsible for the Lamb shift term $\hat{H}_\mathrm{Lamb}$ in Eq.~\eqref{master_eq}. Weighting factor $\mathcal{W}_{\mu}(\omega)$ is related to $A_{\mu}^{>}(\omega)$ through
the Hilbert transform
\begin{equation}
\label{W_appendix}
\mathcal{W}_{\mu}(-E) := J_{\mu}^{2}\,
\mathrm{Pr}\int_{-\infty}^{\infty}d\omega'
\frac{A_{\mu}^{>}(\omega')}{{E}/\hbar-\omega'},
\end{equation}
$\mathcal{W}_{\mu}(E)$ can be written in terms of the susceptibility by making use of the fluctuation-dissipation (FD) relations for $A_{\mu\nu}^{>}(\omega)$:\cite{endnote:4}
\begin{equation}
\label{Amu_FD}
A_{\mu\nu}^{>}(\omega)
= \delta_{\mu\nu}\frac{\mathrm{Im}\chi_{\mu}(\omega)}{\pi(g\mu_B)^2}
\left[1+f_{B}(\hbar\omega/k_B T)\right] .
\end{equation}
By substituting Eq.~(\ref{Amu_FD}) in Eq.~(\ref{W_appendix}),
noting the property
\begin{math}
\mathrm{Im}\chi_\mu(-\omega)=-\mathrm{Im}\chi_\mu(\omega),
\end{math}
and defining the normalized weighting factor as
\begin{equation*}
W_\mu(E) := -\mathcal{W}_\mu(E)\,\frac{(g\mu_B)^2}{\chi_\mu(0)J_\mu^2},
\end{equation*}
we obtain Eq.~(\ref{Wmu}) of the main text.

The second and third line of Eq.~(\ref{LindbladRk}) describe the dissipative effects of the Kondo impurity on the probe spin and constitute dissipation term $\calR[\hat\rho_\mathrm{p}^{(\mathrm{I})}]$ in Eq.~\eqref{master_eq}. A tedious but straightforward inspection of the diagonal part of $\calR[\hat\rho_\mathrm{p}^{(\mathrm{I})}]$ is given by
\begin{multline}
\label{diagelem}
\langle m|\mathcal{R}[\hat{\rho}_{\mathrm{p}}^{(\mathrm{I})}]|m\rangle
= \sum_{n(\ne m)}\bigg(\Gamma_{n\rightarrow m}\langle n|\hat{\rho}_{\mathrm{p}}^{(\mathrm{I})}(t)|n\rangle \\{}
-\Gamma_{m\rightarrow n}\langle m|\hat{\rho}_{\mathrm{p}}^{(\mathrm{I})}(t)|m\rangle\bigg),
\end{multline}
where the relaxation rate $\Gamma_{n\rightarrow m}$ from state $|n\rangle$
to state $|m\rangle$ is given by
\begin{multline}
\hbar\Gamma_{n\rightarrow m} =
2\sum_{\mu}\frac{J_{\mu}^{2}}{(g\mu_B)^2}|S_{nm}^{\mu}|^{2}\mathrm{Im}\chi_{\mu}\left(\frac{E_{nm}}{\hbar}\right)\\{}\times
\left[1+f_{B}\left(\frac{E_{nm}}{k_{B}T}\right)\right],
\end{multline}
where we have used~\eqref{Amu_FD}.
Similarly, the off-diagonal part of $\calR[\hat\rho_\mathrm{p}^{(\mathrm{I})}]$ is given by
\begin{equation}
\label{offdiagelem}
\begin{split}
\langle m|\mathcal{R}[\hat{\rho}_{\mathrm{p}}^{(\mathrm{I})}(t)]|n\rangle
= &-\frac{1}{2}\gamma_{mn}
\langle m|\hat{\rho}_{\mathrm{p}}^{(\mathrm{I})}(t)|n\rangle \\&
-\frac{1}{2}\sum_{l(\ne m)}\Gamma_{m\rightarrow l}
\langle m|\hat{\rho}_{\mathrm{p}}^{(\mathrm{I})}(t)|n\rangle \\&
-\frac{1}{2}\sum_{l(\ne n)}\Gamma_{n\rightarrow l}
\langle m|\hat{\rho}_{\mathrm{p}}^{(\mathrm{I})}(t)|n\rangle
\end{split}
\end{equation}
where
\begin{equation}
\hbar\gamma_{mn}
= \frac{T/T_{K}}{\pi k_{B}T_{K}}\sum_{\mu}2J_{\mu}^{2}(S_{mm}^{\mu}-S_{nn}^{\mu})^2 .
\end{equation}
These expressions allow us to estimate the relaxation time $T_1$ and the dephasing time $T_2$ within the two-level approximation [see also the Bloch equation in Eq.~\eqref{DriveME}] and in the low-temperature limit ($k_BT\ll hf_0\ll k_B T_K$) by
\begin{multline}
2\pi/T_1 = |\Gamma_{1\to 0}| \approx
2\left(\frac{hf_{0}}{k_{B}T_K}\right)
\sum_{\mu}\frac{J_{\mu}^{2}/\hbar}{\pi k_{B}T_{K}}|S_{10}^{\mu}|^{2}
\end{multline}
and
\begin{align}
2\pi/T_2
&= |(\Gamma_{0\to 1} + \Gamma_{1\to 0})/2 + \gamma_{01}/2| \nonumber \\
& \approx |\Gamma_{1\to 0}/2+\gamma_{01}/2| \nonumber \\
& \approx \sum_{\mu}
\frac{J_{\mu}^{2}/\hbar}{\pi k_{B}T_{K}}\left\{ |S_{10}^{\mu}|^{2}\frac{hf_{0}}{k_{B}T_{K}}\right. \\{}&\qquad\qquad\quad
+ \left.
(S_{11}^{\mu}-S_{00}^{\mu})^2\frac{k_{B}T}{k_{B}T_{K}}\right\} .
\label{T2inv_aprx}
\end{align}
Based on the ESR line shape in Eq.~\eqref{If} [or, equivalently, Eq.~\eqref{P1}], one can further estimate the line width $\Delta{f}$ [we define it as the full width at half maximum (FWHM)] by
\begin{equation}
h\Delta{f} \approx
2\sqrt{\left(\sum_{\mu}
  \frac{J_{\mu}^{2}|S_{10}^{\mu}|^{2}}{\pi\hbar\,k_{B}T_{K}}\times
  \frac{h f_0}{k_B T_K}\right)^{2}
  + \frac{1}{2}(\hbar\Omega)^{2}}.
\end{equation}
It has values in the sub-MHz range or less.
Note that the ESR line width is mainly determined by the instrumental factors such as the strong driving RF field and tip-sample vibrations \cite{2015_Science_Baumann}, which is typically in the MHz range.

\section{Alternative effective spin Hamiltonian\label{Sec Appendix C}}

In the main text, the probe spin is described through the spin Hamiltonian (\ref{EffectSpinHam}). By restricting ourselves to impurities with $C_{4v}$ symmetry (see Table~\ref{Table effective parameters}), we consider here a more fundamental model which takes into accounts explicitly the orbital degrees of freedom. 

\begin{figure}
\begin{centering}
\includegraphics[width=8.6cm]{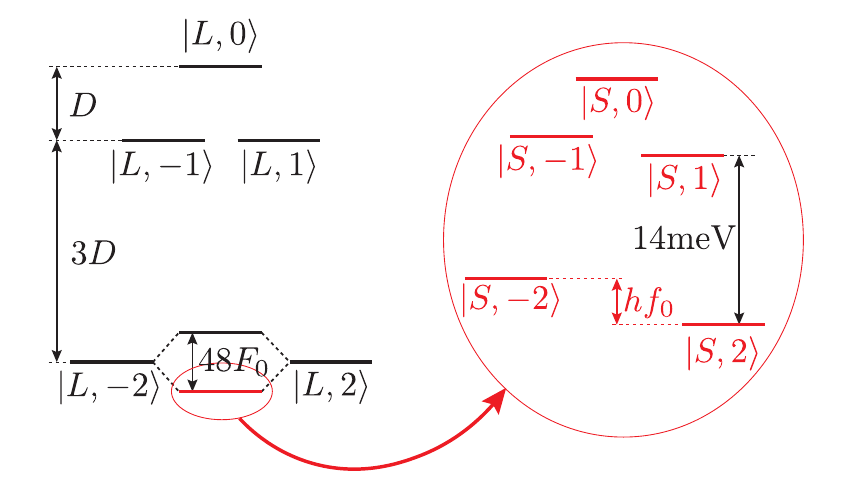}
\par\end{centering}
\caption{Energy level structure of an Fe probe spin with $C_{4v}$ local symmetry. Left panel: Structure of the five $L=2$ orbital
levels. The two lowest orbitals are further mixed by the $F_{0}$ crystal
field term appearing in Eq.~(\ref{CFHamBaumann}).
Right panel: Detailed structure of the low-energy subspace, corresponding to five $S=2$ sub-levels. \label{FIG7 FeEnSpec}}
\end{figure}

A standard derivation of the effective spin Hamiltonian~(\ref{EffectSpinHam}) treats the spin-orbit coupling (SOC) as a perturbation, compared to the orbital level splittings. This assumption is usually well justified for $4f$ electrons \citep{1950_Proc_Phys_Soc_A_Pryce,1999_BOOK_Boca}, but might be violated for adatoms with $3d$ electrons \citep{2012_BOOK_Abragam,1999_BOOK_Boca}. In this case, the crystal field Hamiltonian $\hat{H}_{\mathrm{CF}}=\sum_{i}\hat{H}_{\mathrm{CF}}^{(i)}$ is decomposed into various terms with decreasing strengths $|\hat{H}_{\mathrm{CF}}^{(1)}|>|\hat{H}_{\mathrm{CF}}^{(2)}|>..>|\hat{H}_{\mathrm{CF}}^{(k)}|$, and the spin-orbit term $\hat{H}_{\mathrm{SOC}}$ affects the system at some intermediate level $|\hat{H}_{\mathrm{CF}}^{(1)}|>|\hat{H}_{\mathrm{SOC}}|>|\hat{H}_{\mathrm{CF}}^{(k)}|$, thus a treatment taking into account explicitly the crystal-field is more appropriate. Among interesting impurities falling into this regime is Fe on MgO/Ag(100), which was successfully exploited as probe spin in ESR-STM experiments \cite{2015_Science_Baumann,2017_Nat_Nano_Choi}. Its energy level structure is schematically shown in Fig.~\ref{FIG7 FeEnSpec}.

In case the total orbital $L$ and total spin $S$ angular momentums of the probe atom's electrons are still approximately good quantum numbers, the crystal field Hamiltonian $\hat{H}_{\mathrm{CF}}$ can be constructed by the method of operator equivalence \citep{1952_ProcPhysSocA_Stevens,2012_BOOK_Abragam}. For impurities with $C_{4v}$ or $C_{\infty v}$ local symmetry, the general form of $\hat{H}_{\mathrm{CF}}$ reads 
\citep{2015_Science_Baumann}:
\begin{equation}
\hat{H}_{\mathrm{CF}}=D\hat{L}_{\mathrm{p}}^{z}\hat{L}_{\mathrm{p}}^{z}+E_{0}\left(\hat{L}_{\mathrm{p}}^{z}\right)^{4}+F_{0}\left(\hat{L}_{+}^{4}+\hat{L}_{-}^{4}\right).\label{CFHamBaumann}
\end{equation}
The other two terms entering the total Hamiltonian, $\hat{H}_{\mathrm{atom}}=\hat{H}_{\mathrm{CF}}+\hat{H}_{\mathrm{SOC}}+\hat{H}_{\mathrm{Z}}$, are the spin-orbit coupling:
\begin{equation}
\hat{H}_{\mathrm{SOC}}=\lambda_{z}\hat{L}_{\mathrm{p}}^{z}\hat{S}_{\mathrm{p}}^{z}+\frac{\lambda_{\perp}}{2}\left(\hat{L}_{+}\hat{S}_{-}+\hat{L}_{-}\hat{S}_{+}\right),
\end{equation}
and the Zeeman splitting:
\begin{equation}
\hat{H}_{\mathrm{Z}}=g\mu_{B}\mathbf{B}\cdot\hat{\mathbf{S}}_{\mathrm{p}}+g_{L}\mu_{B}\mathbf{B}\cdot\hat{\mathbf{L}}_{\mathrm{p}}.
\end{equation}
Considering Fe on MgO/Ag(100), theoretical and experimental studies
give the values $D=-433\mathrm{meV}$, $E_{0}=0$, $F_{0}=2.19\mathrm{meV}$, $\lambda_{z}=\lambda_{\perp}=-12.6\mathrm{meV}$, $g=2$ and $g_{L}=1$ \citep{2015_PRL_Baumann,2015_Science_Baumann}. As shown in Fig.~\ref{FIG7 FeEnSpec}, the $F_{0}$ term splits $|L,\pm2\rangle$ states by $48F_{0}$, which is of the same order as the spin-orbit coupling term $2\lambda_{z,\perp}S_{\mathrm{p}}^{z}$. Thus, it is not a good approximation to treat the SOC as a perturbation for the two lowest eigenstates of $\hat{H}_{\mathrm{CF}}$.

\begin{figure}
\begin{centering}
\includegraphics[width=8.6cm]{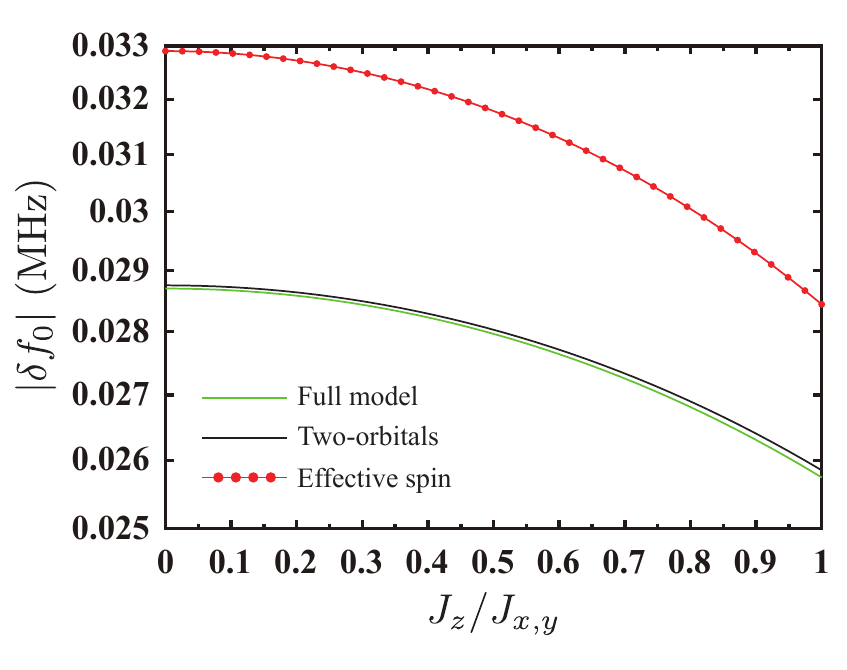}
\par\end{centering}
\caption{ESR frequency shift as a function of the ratio $J_z/J_{x,y}$, for an Fe probe spin on MgO [cf. the black dashed curve in Fig. \ref{FIG 2 typical adatom shifts}]. The green and black solid curves are calculated from the full model Eq.~(\ref{ProbHamPartite}) and the two-orbitals model Eq.~(\ref{ProbEffHam}), respectively. The red circles are obtained from the effective spin Hamiltonian Eq.~(\ref{EffectSpinHam}), with parameters listed in Table~\ref{Table effective parameters}. Other parameters: $L=S=2$,
$D=-433~\mathrm{meV}$, $F_{0}=2.19~\mathrm{meV}$, $\lambda_{z}=\lambda_{\perp}=-12.6~\mathrm{meV}$, $B_{z}=0.2~\mathrm{T}$, $B_x=6~\mathrm{T}$, $T=0.5~\mathrm{K}$, $k_B T_K=1~\mathrm{meV}$, and $J_x=J_y=1~\mathrm{\mu eV}$.\label{TwoOrbitalsModel}}
\end{figure}

Despite this apparent difficulty, in Fig.~\ref{TwoOrbitalsModel} we compare the ESR frequency shift $\delta f_0$ computed from $\hat{H}_\mathrm{atom}$ or the effective spin Hamiltonian Eq.~(\ref{EffectSpinHam}). Besides a relatively small difference (which, in principle, can be compensated by small adjustments of the effective parameters in Table~\ref{Table effective parameters}) the two models are in agreement in predicting a small value $\delta f_0 \lesssim 30$~Hz, weakly sensitive to the anisotropy of the exchange interaction.

We can also derive an alternative effective model, by restricting ourselves to the subspace spanned by $|L,\pm2\rangle$ and eliminating direct transitions to the higher orbital levels. To this end, we make the following
partition of the total Hamiltonian:
\begin{equation}
\hat{H}_{\mathrm{atom}}=\hat{H}_{0}+\hat{H}_{1}+\hat{H}_{2},\label{ProbHamPartite}
\end{equation}
where the unperturbed Hamiltonian is:
\begin{equation}
\hat{H}_{0}=D(\hat{L}_{\mathrm{p}}^{z})^{2}+F_{0}(\hat{L}_{+}^{4}+\hat{L}_{-}^{4})+\lambda_{z}\hat{L}_{\mathrm{p}}^{z}\hat{S}_{\mathrm{p}}^{z},
\end{equation}
which includes here both crystal-fled and spin-orbit coupling terms. The two perturbations are given by:
\begin{equation}
\hat{H}_{1}=\mu_{B}\hat{L}_{\mathrm{p}}^{x}B_{x}+\frac{1}{2}\lambda_{\perp}(\hat{L}_{+}\hat{S}_{-}+\hat{L}_{-}\hat{S}_{+}),
\end{equation}
as well as the remaining of the Zeeman term (assuming $\mathbf{B}$ in the $x$-$z$ plane)
\begin{equation}
\hat{H}_{2}=\mu_{B}B_{z}(\hat{L}_{\mathrm{p}}^{z}+2\hat{S}_{\mathrm{p}}^{z})+2\mu_{B}B_{x}\hat{S}_{\mathrm{p}}^{x}.
\end{equation}
We obtain the effective Hamiltonian by applying a Schrieffer-Wolff transformation \citep{1966_PR_Schrieffer}:
\begin{equation}
\hat{H}'_\mathrm{probe}\approx\hat{H}_{0}+\frac{1}{2}[\hat{H}_{1},\hat{\mathcal{S}}]+e^{-\hat{\mathcal{S}}}\hat{H}_{2}e^{\hat{\mathcal{S}}},\label{TransHam}
\end{equation}
where $\hat{\mathcal{S}}$ satisfies $\hat{\mathcal{S}}=-\hat{\mathcal{S}}^{\dagger}$ and $[\hat{\mathcal{S}},\hat{H}_{0}]=\hat{H}_{1}$, i.e., it is chosen to eliminate $\hat{H}_{1}$ to the lowest order. Since $\hat{H}_{2}$ is already small, we neglect in the last term of Eq.~(\ref{TransHam}) the corrections induced by the unitary transformation. Furthermore, $\hat{\mathcal{S}}$ involves transitions between the $|L,\pm 2\rangle$ and $|L,\pm 1\rangle$ subspaces, and we can set the energy denominators equal to $3D$. As illustrated in Fig.~\ref{TwoOrbitalsModel}, this approximation neglects corrections to the eigenenergies induces by $F_0$ and and $\lambda_z$ which, however, are much smaller than $D$. Finally, after dropping a small shift equal to $(\mu_{B}B_{x})^{2}/(3D)$, we obtain:
\begin{align}
\hat{H}'_{\mathrm{probe}}= &4D-6D^\prime_{s}+24F_{0}\hat{\tau}^{x}+D^\prime_{s}(\hat{S}_{\mathrm{p}}^{z})^{2}+\lambda_{z}'\hat{S}_{\mathrm{p}}^{z}\hat{\tau}^{z}\nonumber \\
&+2\mu_{B}B_{z}(\hat{S}_{\mathrm{p}}^{z}+\hat{\tau}^{z})+\mu_{B}g_{x}B_{x}\hat{S}_{\mathrm{p}}^{x} ,\label{ProbEffHam}
\end{align}
where $\hat{\tau}^{\mu}$ are Pauli matrices of a pseudo spin-1/2
particle spanned by the two lowest orbital states: 
\begin{align}
\hat{\tau}^{x}=|L,2\rangle\langle L,-2|+|L,-2\rangle\langle L,2|,\nonumber \\
\hat{\tau}^{z}=|L,2\rangle\langle L,2|-|L,-2\rangle\langle L,-2|,
\end{align}
and the coefficients in Eq. (\ref{ProbEffHam}) are given by:
\begin{equation}
D^\prime_{s}=\lambda_{z}'-2\lambda_{z}=-\frac{\lambda_{\perp}^{2}}{3D},\text{ }g_{x}=2\left(1+\frac{\lambda_{\perp}}{3D}\right).\label{UniaxialEffPara}
\end{equation}
The effective spin Hamiltonian Eq. (\ref{ProbEffHam}) is valid provided that $F_{0},|\lambda_{z,\perp}|\ll D$ are satisfied, and is applicable to adatoms with $C_{4v}$ or $C_{\infty v}$ local symmetry. In Fig.~\ref{TwoOrbitalsModel}, we see that the ESR frequency shift $\delta f_0$ obtained from $\hat{H}^\prime_\mathrm{probe}$ is in excellent agreement with the full model.

In closing, we note that the parameters of the two-orbitals effective Hamiltonian also allow for a transparent description of ESR-STM setups. For example, according to Ref.~\onlinecite{2015_Science_Baumann}, the ESR drive modifies the local electrostatic environment in a time-dependent way, yielding a cosine modulation of $F_0$. Such coupling appears explicitly in Eq.~(\ref{ProbEffHam}) and, for a given strength of the $F_0$ modulation, it would be possible to compute in a rather straightforward way the Rabi frequency of the ESR transition. Instead, the ESR drive enters the effective parameter of Eq.~(\ref{EffectSpinHam}) in a more complicated manner.

\section*{Acknowledgments}

Y.F.\ acknowledges support from NSFC (Grant No. 12005011).
S.C.\ acknowledges support from NSFC (Grant No. 11974040) and
NSAF (Grant No. U1930402).
M.-S.C.\ has been supported by the National Research Function (NRF) of Korea
(Grant Nos. 2017R1E1A1A03070681 and 2018R1A4A1024157) and by the Ministry of
Education through the BK21 Four program.

\bibliographystyle{apsrev}
\bibliography{References}

\begin{thebibliography}{80}
\bibitem[{end({\natexlab{a}})}]{endnote:3}
\bibinfo{note}{The additional `-' sign in Eq.~\eqref{DynSusp} comes from the
  same sign in the implied perturbation term, $-\sum_\mu
  g\mu_BB_\mu\hat{S}_K^\mu$, due to an external magnetic field that is
  consistent with Eq.~\eqref{chi_Lorenzian}.}
\bibitem[{end({\natexlab{d}})}]{endnote:4}
\bibinfo{note}{A fluctuation-dissipation relation usually appears with a `-'
  sign. It is missing here because of the implied perturbation term, $-\sum_\mu
  g\mu_BB_\mu\hat{S}_K^\mu$; see also Eqs.~\eqref{DynSusp} and
  \eqref{chi_Lorenzian}.}
\end{thebibliography}

\end{document}